\documentclass[twocolumn,showpacs,preprintnumbers]{revtex4}
\usepackage{amsfonts}
\usepackage{amsmath}
\usepackage{graphicx}
\usepackage{dcolumn}
\usepackage{bm}

\input{tcilatex}

\begin{document}

\preprint{}
\title{High energy cosmic rays, gamma rays and neutrinos from AGN}
\author{Yukio Tomozawa}
\affiliation{Michigan Center for Theoretical Physics, Randall Laboratory of Physics,
University of Michigan, Ann Arbor, MI. 48109-1040}
\date{\today }

\begin{abstract}
The author reviews a model for the emission of high energy cosmic rays,
gamma-rays and neutrinos from AGN (Active Galactic Nuclei) that he has
proposed since 1985. Further discussion of the knee energy phenomenon of the
cosmic ray energy spectrum requires the existence of a heavy particle with
mass in the knee energy range. A possible method of detecting such a
particle in the Pierre Auger Project is suggested. Also presented is a
relation between the spectra of neutrinos and gamma-rays emitted from AGN.
This relation can be tested by high energy neutrino detectors such as
ICECUBE, the Mediterranean Sea Detector and possibly by the Pierre Auger
Project.
\end{abstract}

\maketitle

\section{Introduction}

High energy gamma rays from AGN have been reported by\ the EGRET detector at
the Compton Observatory \cite{compton} and by Cerenkov Detectors\cite%
{cerenkov}. More recently a possible correlation between high energy cosmic
rays and the distribution of AGN locations has been reported\cite{auger}. In
this respect, the author had proposed a model for the emission of high
energy particles from AGN since 1985, and the implications of the knee
energy in high energy cosmic ray spectrum had been clarified. In this
report, the author summarizes the proposed model and further explores the
implication of the knee energy. As an additional prediction of the model,
the ratio of neutrinos and gamma rays from AGN is computed.

\section{Summary of the model}

In a series of articles\cite{cr1}-\cite{cr9}, the author has presented a
model for the emission of high energy particles from AGN. The following is a
summary of the model.

1) Quantum effects on gravity yield repulsive forces at short distances\cite%
{cr1},\cite{cr3}.

2) The collapse of black holes results in explosive bounce back motion with
the emission of high energy particles.

3) Consideration of the Penrose diagram eliminates the horizon problem for
black holes\cite{cr4}. Black holes are not black anymore.

4) The knee energy for high energy cosmic rays can be understood as a split
between a radiation-dominated region and a matter dominated region, not
unlike that in the expansion of the universe. (See page 10 of the lecture
notes\cite{cr1}-\cite{cr3}.)

5) Neutrinos and gamma rays as well as cosmic rays should have the same
spectral index from each AGN. They should have the knee energy phenomenon, a
break in the energy spectral index, similar to that for the cosmic ray
energy spectrum.

6) The recent announcement by Hawking rescinding an earlier claim about
information paradox\cite{hawking} is consistent with this model.

The discussion of the knee energy in the model is recapitulated in the next
section, both for the sake of the subject itself, and to introduce the
formulation of the model in order to calculate the ratio of neutrino flux
and gamma ray flux from AGN. This discussion yields the existence of a new
mass scale at the knee energy, which is presented in the subsequent section.
The computation of the neutrino flux and gamma ray flux follows.

\section{The high energy cosmic ray knee energy}

There is a break in the slope of the cosmic ray energy spectrum at around a
few PeV (=10$^{15}$eV). This value is called the knee energy. This
phenomenon was explained as a change of expansion rate in the model\cite{cr1}%
-\cite{cr3}. The number of particles of type X with spin $s$ emitted with
energy E is given by%
\begin{equation}
f_{X}(E)=\frac{2s+1}{2\pi ^{2}}\int \eta _{X}(E/kT)\frac{E^{2}V_{S}dt}{%
e^{E/kT-\mu /kT}\pm 1},  \label{eq1}
\end{equation}%
where $V_{S}$ is the effective volume around the surface of the system with
temperature $T$\ that emits particles, \ $\eta _{X}(E/kT)$ is the emissivity
and $\mu $ is the chemical potential. The $\pm $ sign in the denominator is
for fermions/bosons. It is reasonable to assume that the emissivity for
black holes with a repulsive potential core is close to unity, since
absorption is perfect for black holes and reflection is sure to follow from
the repulsive force:%
\begin{equation}
\eta _{X}(E/kT)\approx 1.
\end{equation}%
With the assumption%
\begin{equation}
V_{S}=\frac{4\pi a}{(kT)^{3}}
\end{equation}%
and the expansion rate%
\begin{equation}
t=bR^{\alpha }  \label{rate}
\end{equation}%
and%
\begin{equation}
R=\frac{d}{kT},
\end{equation}%
where $a$, $b$ and $d$ are constants, one can compute the number of
particles, $f_{X}(E)$, in Eq. (\ref{eq1}),%
\begin{equation}
f_{X}(E)=\frac{A_{X,\alpha }}{E^{\alpha +1}},
\end{equation}%
where%
\begin{equation}
A_{X,\alpha }=\frac{2(2s+1)ab\alpha d^{\alpha }}{\pi }\int_{0}^{\infty }%
\frac{x^{\alpha +2}dx}{e^{x-\mu _{0}}\pm 1}
\end{equation}%
and%
\begin{equation}
\mu _{0}=\mu /kT,\text{ \ \ }x=E/kT.
\end{equation}%
From the expansion rate in cosmology, the exponent $\alpha $ in Eq. (\ref%
{rate}) can be estimated as%
\begin{equation}
\alpha =2\text{ }at\text{ }high\text{ }temperature\text{ }or\text{ }radiation%
\text{ }do\min ated\text{ }era
\end{equation}%
and%
\begin{equation}
\alpha =3/2\text{ }at\text{ }low\text{ }temperature\text{ }or\text{ }matter%
\text{ }do\min ated\text{ }era.
\end{equation}%
For the latter, the exponent varies depending on the assumed mixture of
radiation and matter. Note that all relativistic particles behave like
radiation as far as the relationship between pressure and energy density is
concerned. (i.e., $p=\rho /3$ for radiation as well as for relativistic
particles.)

For high temperature or high energy ($\alpha =2$), the energy spectrum
behaves as%
\begin{equation}
f_{X}(E)\approx 1/E^{3}
\end{equation}%
and for low temperature or low energy ($\alpha =1.5$), the energy spectrum
behaves as%
\begin{equation}
f_{X}(E)\approx 1/E^{2.5}.
\end{equation}%
This is consistent with the observed cosmic ray energy spectrum. It explains
the existence of the knee energy in the cosmic energy energy spectrum. How
about its magnitude? That is the subject of the next section.

\section{The existence of a new mass scale}

At the knee energy of a few PeV or at the corresponding temperature, all
particles of rest mass below the GeV or TeV scale are moving with
relativistic speed, so they contribute as radiation-like or they give the
relationship, $p=\rho /3$. Then, one cannot get a knee energy at the PeV
scale in the model of this article. They would give a knee energy in their
rest mass range. In order to create a knee energy in the cosmic ray spectrum
at a few PeV, one has to assume the existence of particles of a few PeV mass
in black holes. Or equivalently there has to exist a mass scale that creates
significant amounts of particles of a few PeV inside black holes, so that a
significant departure from the radiation-like relationship, $p=\rho /3$, is
established. This can be viewed as observational evidence for a new mass
scale at a few Pev. It is, then, important to establish such a mass scale by
other experimental observations.

An important question is what is the least massive particle at this new mass
scale, so that it can be observed as a stable or at least a quasi-stable
particle. Let us consider some possibilities.

A. The lowest mass is less than a few TeV. This particle can be produced at
LHC. This is a lucky case for the LHC experimental groups. However, it is
not clear whether one can produce a cosmic ray knee energy at the PeV range
by unstable particles at these mass ranges.

B. The lowest mass is larger than a few TeV. This is an unlucky case for
LHC. The Pierre Auger Project would be the only way to detect such
particles. In the analysis of secondary shower particles, one can inquire
whether various combinations of subsystems can make a bump at a particular
mass value. That can be applied to a search for stable or unstable
particles. One can make a search for masses with a fraction of a PeV for
production in the atmosphere or up to the Pev mass scale for stable
particles produced in black holes. These are conceivable for the Pierre
Auger Project.

If any of the stable particles discussed above is weakly interacting, it
becomes a candidate for dark matter. For a discussion of a PeV-scale
supersymmetric theory, see\ e.g., reference\cite{pevss}.

If one gives a name for particles at this PeV mass scale, a designation
based on the word for knee may be appropriate. If one uses a latin word,
genu (or ginocchio for italian), it may become genon. Since this could be
confused with a biological usage, one may call it simply kneeon\cite%
{alternative}.

\section{Ratio of neutrino flux and gamma ray flux}

The flux of gamma rays and neutrinos in this model can be written as ($%
\lambda =\alpha +1$)%
\begin{equation}
f_{X}(E)=\frac{A_{X,\lambda -1}}{E^{\lambda }}
\end{equation}%
where%
\begin{equation}
A_{\gamma ,\lambda -1}=\frac{2K_{\lambda }}{\pi }\int_{0}^{\infty }\frac{%
x^{\lambda +1}}{e^{x}-1}dx
\end{equation}%
for gamma rays and%
\begin{equation}
A_{\nu ,\lambda -1}=\frac{3K_{\lambda }}{\pi }\int_{0}^{\infty }\frac{%
x^{\lambda +1}}{e^{x}+1}dx
\end{equation}%
for neutrinos, where%
\begin{equation}
K_{\lambda }=2ab(\lambda -1)d^{\lambda -1}
\end{equation}%
Here, the chemical potential for gamma rays is zero, and the same is assumed
for mass zero neutrinos. Then, one gets%
\begin{equation}
\frac{A_{\nu ,\lambda -1}}{A_{\gamma ,\lambda -1}}=\frac{3\sum_{n=1}^{\infty
}(-1)^{n}\frac{\Gamma (\lambda +2)}{n^{\lambda +2}}}{2\sum_{n=1}^{\infty }%
\frac{\Gamma (\lambda +2)}{n^{\lambda +2}}}=\frac{3}{2}(1-\frac{1}{%
2^{\lambda +1}})
\end{equation}%
This ratio is given as%
\begin{equation}
\frac{3}{2}(7/8,\text{ }0.912,\text{ }15/16)\text{ \ \ for \ \ }\lambda =(2,%
\text{ }2.5,\text{ }3)
\end{equation}%
Obviously, the factor 3 for the neutrino flux is for three kinds of
neutrinos and the factor 2 for the gamma rays is for two polarizations. The
three neutrinos have equal intensities and there is no oscillation when they
propagate. The last statement is independent of the massless assumption for
neutrinos.

It is also obvious that gamma rays and neutrinos should have the same
spectral index and that both should have the same knee energy for each AGN.\
As the result of a new mass scale, the knee energy has an approximately
universal value. Each AGN has different physical conditions for
gravitational collapse and subsequent expansion and each can have different
spectral index below the knee energy. However, a spectral index of 3 above
the knee energy is most likely to be universal since it is caused by
radiation domination. Below the knee energy, the spectral index is decided
by the mixing of radiation-dominated component of light particles and a
matter-dominated component of heavy particles. Thus, each AGN can have
different spectral index, as seen in the gamma ray observations\cite{compton}%
,\cite{cerenkov}. More explicitly,%
\begin{equation}
\text{for \ \ }p=\frac{\epsilon }{3}\rho ,
\end{equation}%
with%
\begin{equation}
0\leq \epsilon \leq 1,
\end{equation}%
one gets%
\begin{equation}
\alpha =\frac{3+\epsilon }{2}
\end{equation}%
and 
\begin{equation}
\lambda =\alpha +1=2.5+\epsilon /2.
\end{equation}%
If one includes the curvature term in the Friedman metric, one ends up with
a more complicated form.

\section{Discussion}

The observation of a knee energy at a few PeV implies the existence of a new
mass scale in nature in this model. It may be characteristic of \ the
appearance of new physics, a conclusion already indicated from the
observation of high energy cosmic rays with a knee energy. It suggests to
find such evidence from other experimental observation, most likely from a
detailed analysis of the Pierre Auger Project data.

A further important point of this model is that the properties of black
holes have been completely changed, a revolutionary change. The highest
energy cosmic rays observed on Earth may be coming from black holes, an
irony considering the classical description of black holes, \ which tells us
that nothing comes out of them. The Hawking radiation emitted by massive
black holes is miniscule, since the Hawking temperature is inversely
proportional to the black hole mass. High energy gamma rays as well as high
energy neutrinos are coming out of them, too.

Jets coming from AGN can be understood from this model. They are low energy
components, which were accummulated by the magnetic field in the AGN.

In this modified form of black holes, a lot of stuff is coming out. As a
result, the mass of a black hole can be reduced by the emission process. In
the old picture, the mass of a black hole can only increase. This creates a
new scenario for the development of AGN or black holes. The core of AGN can
become smaller on a historical time scale. This may have something to do
with the observational fact that large massive black holes in AGN are at
larger distances or at a long gone past, while all galaxies in our
neighborhood have core black holes with smaller masses at the center. One
can postulate that an ordinary galaxy with a small central black hole can be
a derivative of an AGN from the old days that has a massive black hole\cite%
{massdistance}.

A black hole can become a neutron star by reducing mass by emission. If the
remnant core of SN 87A is a black hole at the present time, it may become a
neutron star at some later time. That is a possible scenario now.

Finally, I should point out the possibility of the formation of massive
black holes made of dark matter. If they are surrounded by dark matter, then
one does not observe them as visible AGN. In other words, they are invisible
AGN. Then, such objects can emit particles, dark matter particles, gamma
rays, neutrinos and possibly ordinary cosmic rays. They become sourceless
cosmic rays as far as the optical signature is concerned. In such cases, the
only signature is high energy gamma rays and neutrinos. A recent report on
high energy cosmic rays in the northern hemisphere that are not correlated
with known AGN\cite{pysicstoday} may be correlated with invisible AGN. One
has to consider a correlation with gamma ray sources in such a case. One
might ask why such a signature was observed in the northern hemisphere only
so far. There is another observation data where a north-south asymmetry is
observed\cite{center}.

\bigskip

\begin{acknowledgments}
\bigskip The author would like to thank Jean Krisch, James Wells and David
N. Williams for useful discussions.
\end{acknowledgments}


\begin{thebibliography}{99}
\bibitem{alternative} If one objects to the use of this terminology because
of the proximity of its pronunciation to that of the 10 th atom, Neon, one
may use sion or xion from the Chinese word for knee, xi, (pronounced si),
and the Japanese, sitsu, or hizon from the Japanese word for knee, hiza.
(Japanese words have a double structure, one from a Chinese derivative and
the other from Japanese proper.)
\end{thebibliography}
\end{document}